\documentclass{jfm} 

\usepackage{graphicx}
\usepackage{natbib}
\usepackage{dcolumn}
\usepackage{bm}

\ifCUPmtlplainloaded \else
  \checkfont{eurm10}
  \iffontfound
    \IfFileExists{upmath.sty}
      {\typeout{^^JFound AMS Euler Roman fonts on the system,
                   using the 'upmath' package.^^J}%
       \usepackage{upmath}}
      {\typeout{^^JFound AMS Euler Roman fonts on the system, but you
                   dont seem to have the}%
       \typeout{'upmath' package installed. JFM.cls can take advantage
                 of these fonts,^^Jif you use 'upmath' package.^^J}%
      }
  \else
  \fi
\fi

\ifCUPmtlplainloaded \else
  \checkfont{msam10}
  \iffontfound
    \IfFileExists{amssymb.sty}
      {\typeout{^^JFound AMS Symbol fonts on the system, using the
                'amssymb' package.^^J}%
       \usepackage{amssymb}%
       \let\le=\leqslant  
       \let\ge=\geqslant  
      }{}
  \fi
\fi

\ifCUPmtlplainloaded \else
  \IfFileExists{amsbsy.sty}
    {\typeout{^^JFound the 'amsbsy' package on the system, using it.^^J}%
     \usepackage{amsbsy}}
    {\providecommand\boldsymbol[1]{\mbox{\boldmath $##1$}}}
\fi

\providecommand\bnabla{\boldsymbol{\nabla}}
\providecommand\bcdot{\boldsymbol{\cdot}}

\title[Steady base states for Navier-Stokes granular dynamics]{Steady base states for Navier-Stokes granular hydrodynamics with boundary heating and shear}

\author[F. Vega Reyes and J. S. Urbach]%
{F\ls R\ls A\ls N\ls C\ls I\ls S\ls C\ls O\ls \ns  V\ls E\ls G\ls
A\ls \ns  R\ls E\ls Y\ls E\ls S\ls$^1$ \and J\ls E\ls F\ls F\ls R\ls E\ls Y\ls \ns S.\ls \ns  U\ls R\ls B\ls
A\ls C\ls H\ls$^2$ \ns}
\affiliation{$^1$Departamento de F\'{\i}sica, Universidad de
Extremadura, E-06071
Badajoz, Spain. email: fvega@unex.es \\ [\affilskip]
$^2$Department of Physics, Georgetown University,
Washington DC, 20057. email: urbach@physics.georgetown.edu}

\date{\today}

\begin{document}

\maketitle

\begin{abstract}
We study the Navier-Stokes steady states for a low density monodisperse hard sphere granular gas (i.e,  a hard sphere ideal monatomic gas with inelastic interparticle collisions).   We present a classification of the uniform steady states that can arise from shear and temperature (or energy input) applied at the boundaries (parallel walls). We consider both symmetric and asymmetric boundary conditions and find steady states not previously reported, including sheared states with linear temperature profiles.  We provide explicit expressions for the hydrodynamic profiles for all these steady states. Our results are validated by the numerical solution of the Boltzmann kinetic equation for the granular gas obtained by the direct simulation Monte Carlo method, and by molecular dynamics simulations. We discuss the physical origin of the new steady states and derive conditions for the validity of Navier-Stokes hydrodynamics.  
\end{abstract}

\section{Introduction}
\label{intro}

The kinetic theory of non-uniform gases has advanced significantly in
the last few decades \cite*[]{Hilbert,Chapman,Grad,Ernst}. For instance, the Chapman-Enskog method \cite[]{Chapman},
which allows the determination of the transport
coefficients associated with the equations of evolution of the average
fields $n$, $\boldsymbol{v}$ and $T$ (density, flow velocity and granular temperature, respectively), has recently been successfully extended to granular gases, where
energy is not conserved in the collisions \cite[]{Goldhirsch}. However, debate continues
regarding fundamental issues such as the validity, in granular gases, of the assumptions
made in the standard Chapman-Enskog method \cite[]{Goldhirsch}. Nonetheless, this work has produced a series of results that can help explain important features observed in real or computer experiments
\cite[for example, non-equipartition in granular gases mixtures,][]{vicent, barrat, FM02, Wildman}.

The use, in the macroscopic balance equations for a granular gas, of the constitutive equations of the stress tensor and heat flux and the corresponding hydrodynamic coefficients yields a peculiar hydrodynamics from which many interesting transport phenomena arise \cite[]{Goldhirsch}.  
In the hydrodynamic equations a nonuniform temperature profile arises in general from the effects of viscous heating (in cases where there is shear), collisional cooling, and thermal transport. It has been shown that a hard-disk or hard-sphere system that is sheared and heated by two identical infinite parallel walls can have a temperature profile of negative curvature \cite*[when viscous heating exceeds collisional cooling, see works by][]{JS83, NAA99,JStat}, positive curvature (when the reverse is true), or a flat temperature profile \cite[]{Campbell}. This last condition, occurring when viscous heating and collisional cooling exactly balance, is called Uniform Shear Flow (USF), and has been extensively studied \cite[]{Goldhirsch}. In the case of a heated gas (with no shear), temperature profiles with positive curvature \cite[]{Barrat2} have been reported. These  studies considered only symmetric boundary conditions for the granular temperature (for example, both walls at the same temperature). 
Alam, Nott and collaborators \cite[see, for instance][]{AN98, NAA99} and \cite{WJS96} have obtained stability and bifurcation criteria for the base steady flows resulting from symmetric boundary conditions for the temperature, with specific studies on the stability of the uniform shear flow \cite[]{A06}. On the other hand, there are also works where asymmetric boundary conditions are analyzed. In these cases, except for the work by \cite{Fourier}, where a linear temperature profile in an unsheared granular gas is found, the same types of profiles as in the symmetric case are reported: for instance,  \cite{KMS08} derive theoretical criteria for the correction of the temperature jump at the boundary, and \cite{GHW07} study via MD simulations the deviations from NS behaviour in the boundary layer.




In this work we focus on the systematic derivation of all steady profiles and we show that the number of steady states arising from asymmetric boundary conditions for the granular temperature is actually twice the number of types of profiles reported previously. With 'type of profile' we mean profiles that share basic properties like a given temperature curvature, a particular form of the velocity profile and a specific balance between viscous heating, collisional cooling, and the exchange of heat flow with the boundaries. We also describe in this work the basic balance mechanisms of heat flow, viscous heating and collisional cooling that can occur in the steady base states of a granular gas. In effect, we demonstrate that the temperature profile curvature (positive, negative or zero) is determined by an interplay of collisional cooling and viscous heating in the bulk of the fluid and \textit{also} the difference between the granular temperature at the different boundaries of the system. This produces some surprising results, for instance, we find that effects of the temperature boundary conditions may compensate collisional cooling in the bulk even in the absence of shearing and produce a steady temperature profile with negative curvature and we find that it is possible for a sheared granular gas to have a perfectly linear temperature profile and that this happens even though collisional cooling and viscous heating are not balanced (contrary to what happens in the USF). We validate our resulats with simulation data (both Monte Carlo method and molecular dynamics).


The structure of the paper is as follows: In \S~\ref{ktb} we determine the differential equations for the steady temperature and flow velocity profiles that result from the Navier-Stokes equations for the granular gas. In \S~\ref{2D}, we obtain all possible types of unidimensional steady profiles (henceforth, steady base states): in \S~\ref{difeq} we write the differential equations in reduced magnitudes; the resulting form of these equations allows us to define in \S~\ref{redgrad} the relevant reduced (microscopic/hydrodynamic) length and time scales; in \S~\ref{tiposT} we obtain analytical solutions of the steady hydrodynamic profiles and describe the general properties of the resulting temperature profiles; finally, in \S~\ref{clas} we give the classification of the different types of base steady states and flows in granular gases, in the case of constant pressure. We end with \S~\ref{concl}, where we compare our theory with simulation data obtained by the direct simulation Monte Carlo method (DSMC) and molecular dynamics simulations (MD).
 
\section{Continuum balance equations and transport coefficients}
\label{ktb}

With the assumptions of mass and momentum conservation in the collisions and from integration of the Boltzmann equation for a granular gas $\times \{1,m\boldsymbol{v},mv^2/2\}$ the mass, momentum and energy balance equations become \cite[]{Jenkins,Brey,GoldJFM}

\begin{equation}
\frac{Dn}{Dt}=-n\bnabla\bcdot\boldsymbol{u} \: ,
\label{nbal}
\end{equation}

\begin{equation}
\frac{D\boldsymbol{u}}{Dt}=-\frac{1}{mn}\bnabla\bcdot{\mathsfbi{P}} \: ,
\label{Pbal}
\end{equation}

\begin{equation}
\frac{DT}{Dt}+T\zeta=-\frac{2}{dn}\left(\mathsfbi{P}\boldsymbol{:\nabla}\boldsymbol{u}+\bnabla\bcdot{\boldsymbol q}\right) \: ,
\label{Tbal}
\end{equation} where $D/Dt\equiv \partial/\partial t+\boldsymbol{u}\bcdot\bnabla$ is the material derivative, $m$ the particle mass, and $n$ is the particle density, $\boldsymbol{u}$ the flow velocity, $T$ the temperature, $\mathsfbi{P}$ the stress tensor and $\boldsymbol{q}$ the heat flux  \cite[see][and references therein for more details on the definitions and derivations]{Brey,BreyD,Goldhirsch}. The rate at which kinetic energy is lost in the collisions, the cooling rate, is denoted as $\zeta$. The balance in (\ref{Tbal}) between the term coming from viscous heating (the first term on the right hand side) and the collisional cooling (the second term on the left hand side)  will determine the gradient of the heat flux in the steady state ($\partial /\partial t=0$).

 The hydrodynamic description (and thus, our results) is valid only for sufficiently small gradients of $n, \boldsymbol{u}, T$, meaning that the typical microscopic length scale $\lambda$ (the particle mean free path) must be small compared to a typical system macroscopic length scale $L$ (determined by the gradients of $n, \boldsymbol{u}, T$): 
$\lambda/L\ll1$. 

For small enough $\lambda/L$, we may retain in the Chapman-Enskog expansion only terms up to first order in the gradients  \cite[]{Chapman} , and in this case we find the following constitutive relations for the stress tensor and heat flux \cite[]{Brey}

\begin{equation}
\mathsfbi{P}=p \mathsfbi{I}-\eta\left[\boldsymbol\nabla\boldsymbol{u}+\boldsymbol\nabla\boldsymbol{u}^\dagger-\frac{2}{d}(\bnabla\bcdot\boldsymbol{u}) \mathsfbi{I}\right] \: ,
\label{Pconst}
\end{equation} 

\begin{equation}
\boldsymbol{q}=-\kappa\boldsymbol{\nabla} T-\mu\boldsymbol{\nabla} n \: ,
\label{qconst}
\end{equation} where $p=nT$ is the hydrostatic pressure, the coefficients $\eta$ and $\kappa$ are known as the viscosity and heat conductivity respectively, and $\mu$ is a coefficient that only appears when energy is not conserved in the collisions \cite[]{Brey}. Inserting (\ref{Pconst}), (\ref{qconst}) into the balance equations (\ref{nbal})-(\ref{Tbal}) we obtain the Navier-Stokes equations for the granular gas.

The transport coefficients for a low density granular gas of hard spheres have been calculated by \cite{GoldJFM} and \cite{Brey}. We use here the expressions for hard spheres (with the same notation) calculated by \cite{Brey} \cite[see the more recent work,][for more general expressions for a $d$-dimensional system]{BreyD}. From dimensional considerations in (\ref{nbal})-(\ref{Tbal}) with (\ref{Pconst})-(\ref{qconst}), the hydrodynamic Navier-Stokes coefficients have the following form \cite[]{Brey}

\begin{equation}
\eta=\eta_0^*T^{1/2}\: , \quad \kappa=\kappa_0^*T^{1/2} \: ,
\quad \mu=\mu_0^*\frac{T^{3/2}}{n} \: , \quad \zeta=\zeta_0^*\frac{p}{T^{1/2}} \: ,
\end{equation} where $\eta_0^*=\eta^*(\alpha)\eta'_0$, $\kappa_0^*=\kappa'_0\kappa^*(\alpha)$, $\mu_0^*=\kappa'_0\mu^*(\alpha)$ and $\zeta_0^*=\zeta^*(\alpha)/\eta'_0$ are constants. The constants $\eta_0'$ and  $\kappa_0'$ are related respectively to the viscosity, $\eta_0(T)=\eta_0'T^{1/2}$, and thermal conductivity, $\kappa_0(T)=\kappa_0'T^{1/2}$, of an elastic hard disk/sphere gas \cite[]{Chapman, Brey}, with \cite[]{Chapman}:
\begin{eqnarray}
 & & \eta_0'=\frac{2+d}{8}\Gamma(d/2)\pi^{-\frac{d-1}{2}}m^{1/2}\sigma^{-(d-1)}\:, \nonumber \\
 &  &\kappa_0'=\frac{d(d+2)^2}{16(d-1)}\Gamma(d/2)\pi^{-\frac{d-1}{2}}m^{-1/2}\sigma^{-(d-1)} \: .
 \end{eqnarray}

The dimensionless functions $\eta^*(\alpha)$, $\mu^*(\alpha)$, $\kappa^*(\alpha)$ and $\zeta^*(\alpha)$ depend on the coefficient of normal restitution $\alpha$ and in the elastic limit $\eta^*(1), \kappa^*(1)=1$ and $\mu^*(1), \zeta^*(1)=0$. 

Transport coefficients in the quasielastic limit \cite*[]{GoldJFM} and for moderate density granular gases \cite*[]{Jenkins, GD99} and coefficients obtained from expansions from arbitrary reference states \cite*[] {Lutsko,G06} can also be used. Also, alternate hydrodynamic descriptions based on BGK-type kinetic models \cite*[]{VGS07} and coefficients calculated with a more refined Sonine expansions \cite*[]{GSM07, GVM09} may be used.

\subsection{Hydrodynamic steady states}
\label{hydrst}

\begin{figure}
\begin{center}
\includegraphics[height=4.25cm]{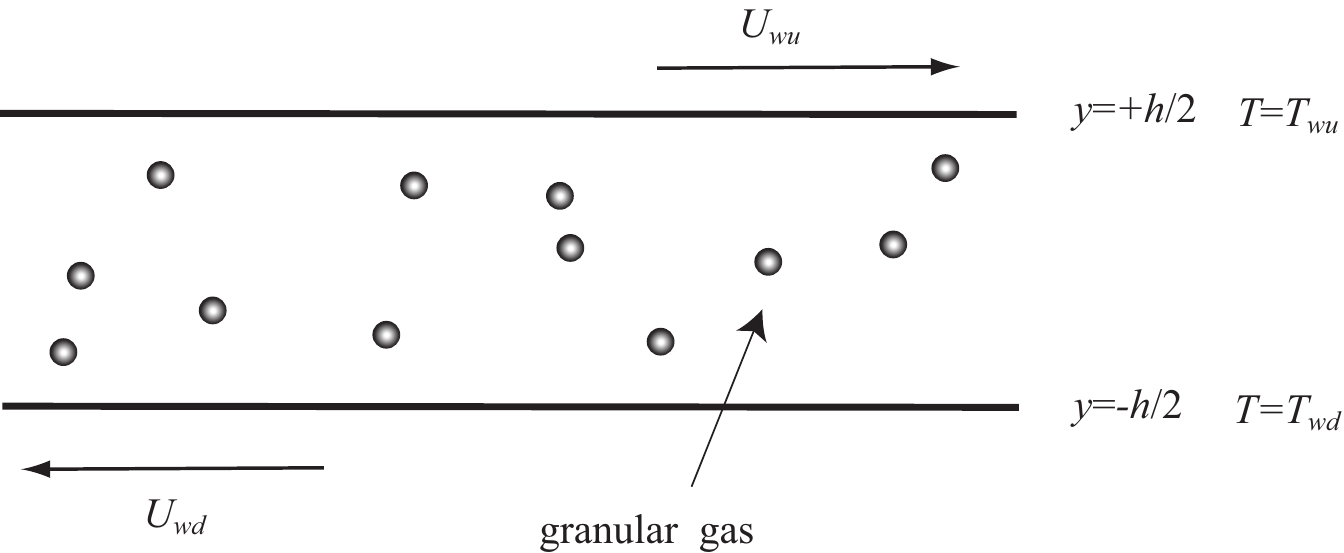}
\end{center}
\caption{Schematic view of the system subject of study. The granular gas is heated and  sheared from two infinite parallel walls.} \label{setup}
\end{figure}

Let us consider first a system with two parallel walls in the planes $y=-h/2$ and $y=+h/2$. The system is infinite in the $x$ and $z$ directions. The walls are kept at temperatures $T_w(y=+h/2)=T_{wu}$ and $T_w(y=-h/2)=T_{wd}$. Additionally, the walls at $y=-h/2$ and $y=+h/2$ may be moving in the $x$ direction, in general with different velocities, so that the granular fluid adjacent to the walls has velocities $\boldsymbol{v}_w(-h/2)=U_{wd}\boldsymbol{e}_x$ and $\boldsymbol{v}_w(+h/2)=U_{wu}\boldsymbol{e}_x$ respectively, where $\boldsymbol{e}_j$ is a unit vector in the $j$ direction. A schematic view of the system is depicted in figure \ref{setup}. 
We will study steady states ($\partial/\partial t=0$) that are uniform in the direction of the flow ($\partial u_x/\partial x=0$), and assume that there is nothing to prevent $\partial p/\partial x=0$ (a gravitational force in that direction, for instance) and we will also assume that the system is uniform in the $z$ direction, $\partial/\partial z=0$. (Thus all conclusions below also apply to a system of sheared two-dimensional inelastic disks, removing coordinate $z$). Under these conditions, our steady states will have the general form:  $\boldsymbol{u}=u(y)\boldsymbol{e}_x$, $T=T(y)$, while $p$ is a constant. 

From (\ref{Pbal}) and (\ref{Tbal}), and taking into account the system symmetries, we derive the Navier-Stokes equations for the steady states,  

\begin{equation}
\sqrt{\frac{T}{T_r}}\frac{\partial u_x}{\partial y}= \frac{\gamma}{\eta_0^*} \: ,
\label{3Dm}
\end{equation}

\begin{equation}
\sqrt{\frac{T}{T_r}}\frac{\partial}{\partial y}\left(\sqrt{\frac{T}{T_r}}\frac{\partial T}{\partial y}\right)=\frac{d}{2}\frac{\zeta^*}{\beta_0^*\eta'_0}\frac{p^2}{T_r}-\frac{\eta_0^*}{\beta_0^*}\frac{T}{T_r}\left(\frac{\partial u_x}{\partial y}\right)^2 \: ,
\label{3DT}
\end{equation} where $\beta_0=\kappa'_0(\kappa^*(\alpha)-\mu^*(\alpha))\equiv\kappa'_0\beta^*(\alpha)$ and we multiplied momentum and energy balance equations by  $1/\sqrt{T_r}$ and $\sqrt{T}/T_r$ respectively.
In (\ref{3Dm}) we have integrated once over $y$, with $\gamma$ the resulting integration constant, which we call \textit{shear rate}; and in (\ref{3DT}) we have taken into account that for constant pressure  $(T^{3/2}/n)(\partial n/\partial y)=-\sqrt{T}(\partial T/\partial y)$. 

Thus, the hydrodynamic problem needs 4 boundary conditions to be closed (two for each $T$ and $u_x$). We will not specify a relation between the hydrodynamic fields at the boundaries and the boundary conditions themselves (i.e., any functions $\boldsymbol{u}(\boldsymbol{v}_w,T_w)$, $T(\boldsymbol{v}_w,T_w)$ at the boundaries).   In \S\ref{tiposT} we discuss the kinds of physical boundaries that would produce the specific profiles we describe.


\section{Analytic solution of unidimensional steady states with zero pressure gradient }
\label{2D}

\subsection{Reduced differential equations}
\label{difeq}

For the microscopic length and time scales we will use the  mean free path $\lambda$ and collision frequency $\nu$ defined at an arbitrary reference point, $\lambda_r$ and $\nu_r$, determined according to

\begin{equation}
\lambda_r=\Lambda_d( \sqrt{2} n_r \sigma^{(d-1)})^{-1}\: ,
\label{lambda}
\end{equation} 

\begin{equation}
\nu_r=\sqrt{\frac{2T_r}{m}}\frac{n_r\sigma^{d-1}}{\Lambda_d}
\: .
\end{equation}

In (\ref{lambda}), $\Lambda_d=\sqrt{2}\frac{2+d}{8}\Gamma (d/2)\pi^{-\frac{d-1}{2}}$ is a constant that depends on the system's dimension $d$, and $T_r$ and $n_r=p_r/T_r$ indicate respectively the values of temperature and density at an (initially arbitrary) reference point. 

All of the partial derivatives in (\ref{3Dm}) and (\ref{3DT}) have a prefactor $\sqrt{T}$, so that with the use of a scaled variable $l$ such that $\sqrt{T/T_r}\partial/\partial y=\partial/\partial l$, the previously non-linear differential equations  (\ref{3Dm}) and (\ref{3DT}) become linear in the new variable \cite*[a similar change of variable was first used by][for studying the steady heat flow in an gas of elastic spheres]{SBG86}. The use of the extra factor $\sqrt{1/T_r}$ allows us to preserve the dimensionality of the spatial gradients in the new variable. Notice also that the resulting summands in (\ref{3Dm}) and (\ref{3DT}) have dimensions of $\nu_h$ and $m\nu_h^2$ respectively; i.e., they set two characteristic (inverse) hydrodynamic time scales for the gradients of $\boldsymbol{u}$ and $T$. Thus, expressing equations (\ref{3Dm}), (\ref{3DT})  as a function of derivatives in the scaled variable $l$ and dividing them by $\nu_r$ and $m\nu_r^2$ respectively, we obtain the following linear and dimensionless differential equations

\begin{equation}
\frac{\partial \hat{u}_x}{\partial\hat{l}}=\frac{1}{\eta^*(\alpha)}\hat{\gamma}\: ,
\label{ugammaRed}\end{equation}

\begin{equation}
\frac{\partial^2 \hat{T}}{\partial\hat{l}^2}=\hat{\Gamma}(\alpha)\: ,
\label{GammaRed}
\end{equation} where we denote the corresponding dimensionless magnitude of $T$ as $\hat{T}=T/T_r$, $y$ as $\hat y=y/\lambda_r$, $u_x$ as $\hat u_x=u_x/(\lambda_r\nu_r)$ etc.

The parameters $\hat\gamma$ and $\hat\Gamma (\alpha)$ read

\begin{equation}
\hat{\gamma}\equiv\frac{\gamma}{p}\: ,
\label{gr}
\end{equation}

\begin{equation}
\hat{\Gamma}(\alpha)\equiv\phi_d\frac{1}{\beta^*(\alpha)\eta^*(\alpha)}\left[\frac{d}{2}\zeta^*(\alpha)\eta^*(\alpha)-\hat{\gamma}^{2}\right]\: ,
\label{GR}
\end{equation} where $\phi_d=2\frac{d-1}{d(d+2)}$.


The dimensionless magnitude $\hat\Gamma(\alpha)$ represents the reduced collisional cooling rate minus the viscous heating rate (coming from the term proportional to $\hat\gamma^2$), and because of this we will call it the \textit {effective cooling rate}. The equilibrium between collisional cooling and viscous heating determines the curvature of the temperature as a function of the scaled variable $\hat l$, but, as we will see in \S~\ref{tiposT}, \textit{it does not the determine} the curvature of $\hat T(\hat y)$.



\subsection{Reduced time and length scales}
\label{redgrad}

The solution of (\ref{GammaRed}), valid for all values of $\hat\gamma$ and $\hat\Gamma (\alpha)$, in terms of the variable $\hat{l}$, is straightforward:

\begin{equation}
\hat{T}(l)=\frac{1}{2}\hat{\Gamma}(\alpha)\hat{l}^2+A\hat{l}+B \: ,
\label{TL}
\end{equation} where $A$ and $B$ are integration constants that will be determined by the physical boundary conditions associated with the temperature. We may identify in the steady states  three different microscopic/macroscopic length (or time) scales. They result from the dimensionless constants $\hat\gamma$, $\hat\Gamma(\alpha)$ and $A$. Hence, we call them (reduced) 'reference scales'.  

In effect, $\hat\gamma/\eta^*(\alpha)=\partial\hat u/\partial\hat l$ is of the order of $\nu_h/\nu_r$ (or equivalently, of the order of $\lambda_r/L$) and $\hat\Gamma(\alpha)=\partial^2\hat T/\partial\hat l^2$ is of the order of $(\nu_h/\nu_r)^2$ (i.e., of the order of $(\lambda_r/L)^2$). Therefore, from (\ref{ugammaRed}, \ref{GammaRed}), $|\hat{\gamma}|/\eta^*(\alpha)$ and $|\hat{\Gamma}(\alpha)|^{1/2}$ set two constant scales for the two (reduced) spatial gradients in the problem. As the reference point for computing $\nu_r$, we pick the one where the temperature has the lowest value, since for this choice a small $\partial/\partial\hat l$ implies small $\partial/\partial\hat y$ (we recall that  $\partial/\partial y=\sqrt{T_r/T}\partial/\partial l$ ).  
The third reference scale is determined by $A$, which yields the reduced temperature gradient imposed at the boundaries: from (\ref{TL}) it is straightforward that $A=(\hat T(+L/2)-\hat T(-L/2))/L$, where $\pm L/2=\hat l(\hat y=\pm h/2)$; therefore, $A$ is of the order of $\nu_h/\nu_r$. 

Thus, our three reference scales may be defined in the following way: $\tau_s\equiv|\hat\gamma|$ (the scale defined by shearing \footnote{$\eta^*(\alpha)$ has little influence since $\eta^*(\alpha)$ is close to 1 for the whole range of values of the coefficient of restitution we consider in this work \cite[see][]{Brey}.}), $\tau_{cs} \equiv |\hat{\Gamma}(\alpha)|^{1/2}$ (the scale define by collisional cooling and viscous heating balance) and $\tau_T\equiv A$ (the scale defined by the wall temperature difference). Their nature and the way they affect the behaviour of the system is qualitatively different in each case. Two of them ($\tau_s$, $\tau_T$) are free; i.e., they are controlled by the boundary conditions. The remaining reference scale ($\tau_{cs}$) is determined jointly by the nature of the granular gas (its degree of inelasticity, and thus this scale is not free like the other two) and the viscous heating. Once the boundary conditions are given, the three scales take, for the value of $\alpha$ of the granular gas, numerical values. We will pick the maximum among these three values as our definition of Knudsen number, denoted by $\textrm{Kn}$. Due to the existence of the inelastic cooling term, $\textrm{Kn}$ cannot be made arbitrarily low as in gases with elastic collisions. Thus, if we look for a given type of flow in a granular gas, there is a minimum value $\textrm{Kn}_m>0$ below which this type flow is not possible. We analyze this in more detail in \S \ref{concl}.



\subsection{Properties of steady temperature profiles (with or without shear)}
\label{tiposT}

The change of variable we use does not change the sign of the first derivative: $\partial\hat{T}/\partial\hat{l}$ has the same sign as $\partial\hat{T}/\partial\hat{y}$, and also $\partial\hat{T}/\partial\hat{l}=0$ if and only if  $\partial\hat{T}/\partial\hat{y}=0$.  We must analyze more carefully the second derivatives:

\begin{equation}
 \frac{\partial^2\hat T}{\partial\hat{y}^2}= \hat T^{-1/2} \frac{\partial }{\partial \hat l}\left(\hat T^{-1/2}\frac{\partial\hat T}{\partial\hat l}\right)=\hat T^{-2}\left[\hat T\frac{\partial^2\hat T}{\partial\hat l^2}-\frac{1}{2}\left(\frac{\partial\hat T}{\partial\hat l}\right)^2\right]=\hat T^{-2}\Phi(\alpha)\: ,
\label{ecder}
\end{equation} with

\begin{equation}
\Phi(\alpha)\equiv B\hat\Gamma (\alpha)-\frac{1}{2}A^2 \label{Phi}
\end{equation}

Relations (\ref{ecder}, \ref{Phi}) imply that, for finite temperature, the curvature of $\hat T(\hat y)$ is either non-zero $\forall\: \hat l, \hat y$, if $\Phi(\alpha)\neq 0$, or zero $\forall\: \hat l, \hat y$ if $\Phi(\alpha)=0$. Furthermore, the sign of $\Phi(\alpha)$ yields the sign of the curvature of $\hat T(\hat y)$. Thus, we call $\Phi(\alpha)$,  defined in (\ref{Phi}), the \textit{temperature curvature parameter}. As we see, the temperature profile curvature is determined by the cooling and viscous heating rate balance, through $\hat\Gamma (\alpha)$, and \textit{also} the boundary conditions for the temperature, through the constants $A, B$. Thus, because of the temperature dependence of the hydrodynamic coefficients, a net thermal flux can reduce or even reverse the effects of the heating and cooling balance on the temperature profile.

Since $\partial/\partial\hat l\equiv\sqrt{\hat T}\partial/\partial\hat y$, we can write

\begin{equation}
\sqrt{\hat{T}}\frac{\partial\hat l}{\partial\hat y}=1 \quad \Rightarrow \quad \int{\sqrt{\hat T(\hat l)}\,d\hat l} =y-y_0\: ,
\label{derivada}
\end{equation} where $y_0$ is an integration constant, to be determined by boundary conditions.

Integration of (\ref{derivada}) involves the square root of the temperature, and for this reason, simpler expressions result if we first rewrite $\hat T(\hat l)$ as a function of terms of the form $(\hat l -l_0)^2$. The equality (\ref{TL}) may be rewritten, for $\hat\Gamma(\alpha)\neq0$, as

\begin{equation}
\hat{T}(\hat{l})=L_0\left[1\pm l_c^2(\hat{l}-l_0)^2\right]\: .
\label{TUmejor}
\end{equation} 

Given that the granular temperature cannot be negative,  we have the following possible combinations, depending on the sign of $L_0$: for $L_0>0$  both upper and lower signs before $l_c^2$ in (\ref{TUmejor}) are possible, with the lower sign  restricted to the interval  $l_c^2(\hat l - l_0)^2\le1$; and for $L_0<0$, only the lower (negative) sign  is possible, and the solution is restricted to the interval $l_c^2(\hat l - l_0)^2\ge 1$. The sets of constants $\{\hat\Gamma(\alpha),A,B\}$ and $\{L_0, l_c, l_0\}$ are related as follows

\begin{equation}
	L_0=B-\frac{A^2}{2\hat\Gamma(\alpha)}=\frac{\Phi(\alpha)}{\hat\Gamma(\alpha)}\: ,\quad \quad
		\pm l_c^2=\frac{\hat\Gamma(\alpha)^2}{2\Phi(\alpha)}\:,	\quad \quad
		l_0=-\frac{A}{\hat\Gamma(\alpha)} \: .
\label{Lrelacion}
\end{equation}

Notice the special case $\Phi(\alpha)=0$ ($A^2=2\hat{\Gamma}(\alpha) B$), for which the relation (\ref{Lrelacion}) between the sets of constants $\{\hat\Gamma(\alpha),A,B\}$ and $\{L_0, l_c, l_0\}$ is indeterminate. In fact, the case $A^2=2\hat{\Gamma}(\alpha) B$ corresponds necessarily to the less general form

\begin{equation}
\hat{T}(\hat{l})=L_{0l}(\hat{l}-l_{0l})^2\: .
\label{TUlineal}
\end{equation} which is fulfilled if and only if $A^2=2\hat{\Gamma}(\alpha)B$ (i.e., $\Phi(\alpha)=0$), and the relation between the constants is

\begin{equation}
	L_{0l}=\frac{\hat\Gamma(\alpha)}{2} \: ,	\quad\quad
		l_{0l}=l_0=-\frac{ A}{\hat\Gamma(\alpha)} \: .
		\label{Lctes}
\end{equation}

Thus, a linear $\hat{T}(\hat{y})$ profile (because $\Phi(\alpha)=0$) necessarily has the form (\ref{TUlineal}) in the variable $\hat l$. The existence of linear temperature profiles in granular gases heated from the boundaries  has not been reported previously, except for a special case in a non-sheared granular gas \cite[]{Fourier}.  Note also that, since the granular temperature cannot be negative, the linear temperature profile (\ref{TUlineal}) is only possible for $L_{0l}>0$; i.e., $\hat{\Gamma}(\alpha)>0$. Inserting (\ref{TUlineal}) in the integral in (\ref{derivada}) we obtain $\hat{y}=y_0\pm L_{0l}^{1/2}\hat{l}(\hat{l}-2l_{0l})/2$, and using this back in (\ref{TUlineal}) we obtain

\begin{equation}
\hat T(\hat y)=\pm2L_{0l}^{1/2}(\hat y-y_0)+L_{0l}l_{0l}^2 \: ,
\end{equation} where the plus and minus signs stand for $\hat l-l_0>0$ and $\hat l-l_0<0$ respectively. If we use this relation, it is straightforward, with the aid of (\ref{Lctes}) and (\ref{GR}), to find that the slope of the temperature profile fulfills

\begin{eqnarray}
& & \frac{1}{p\sigma^{(d-1)}}\frac{\partial T}{\partial y}=\frac{1}{p\sigma^{(d-1)}}\frac{\partial\hat T}{\partial\hat y}\frac{T_r}{\lambda_r}= 
\label{Tyslope} \\ 
& & \nonumber \pm\left(\frac{32(d-1)\pi^{(d-1)}(1-\alpha^2)}{d(d+2)^2\Gamma (d/2)^2 (\kappa^*(\alpha)-\mu^*(\alpha))} -\left(\frac{8}{d+2}\right)^3\frac{(d-1)\pi^{(d-1)} }{d\Gamma(d/2)^2(\kappa^*(\alpha)-\mu^*(\alpha))}\hat\gamma^2\right)^{1/2}.
\end{eqnarray} Notice that this relation includes, as a particular case ($\hat\gamma=0$), the  linear temperature described previously by \cite{Fourier}.

In the case of a sheared system ($\hat{\gamma}\neq0$), once (\ref{derivada}) is integrated the solution of the velocity profile is obtained using (\ref{ugammaRed})

\begin{equation}
\hat{u}_x=\frac{\hat{\gamma}}{\eta^*(\alpha)} \hat{l} + C\: ,
\label{ul}
\end{equation} where $C$ is a constant of integration, that in the following we set equal to zero by a Galilean transformation.

\subsection{General classification of steady hydrodynamic profiles (with or without shear)}
\label{clas}
We have found the following  steady profiles, depending on the values of $\hat{\Gamma}(\alpha)$, $\Phi(\alpha)$ and the constants $A$ and $B$: 

\begin{itemize}
  \item Viscous heating greater than collisional cooling ($\hat{\Gamma}(\alpha)<0$, only with shear):
  \begin{itemize}
 \item{}Temperature profile $T(y)$ with negative curvature ($\Phi(\alpha)<0$):

\begin{eqnarray}
 \hat{T}(\hat{l})=L_0 \left[1-l_c^2(\hat{l}-l_0)^2\right]\: ,\label{TU<}
\end{eqnarray}

\begin{eqnarray}
& &\hat{y}=y_0+L_0^{1/2}\frac{1}{2 l_c}\left[\sin^{-1}(l_c(\hat{l}-l_0))+l_c(\hat{l}-l_0)(1-c^2(\hat{l}-l_0)^2)^{1/2}\right] . \nonumber
\label{TUy<}
\end{eqnarray}
\end{itemize}

  \item Viscous heating equal to collisional cooling ($\hat{\Gamma}(\alpha)=0$, only with shear).
    
  \begin{itemize}
  \item Temperature profile $T(y)$ with negative curvature, but linear $T(u_x)$ ($\Phi(\alpha)<0$ and $A\neq 0$):
  
   \begin{equation}
 \hat T(\hat l)=A\hat l+B\: , \quad\quad\hat{y}=y_0+ \frac{2}{3A}\left(B+A\hat l\right)^{3/2}\: .
 \label{y=}
 \end{equation}

 \item Uniform shear flow (USF): constant temperature, linear velocity profile ($\Phi(\alpha)=0$ and $A=0$, $B\neq0$):
 
   \begin{equation}
 \hat T(\hat l)=B\: , \quad\quad \hat{y}=y_0+B^{1/2}\hat l\: .
\label{y=alt}
 \end{equation}
 
\end{itemize}
  
  \item Collisional cooling greater than viscous heating ($\hat{\Gamma}(\alpha)>0$, with or without shear).
  
  \begin{itemize}

\item Temperature profile $T(y)$ with negative curvature ($\Phi(\alpha)<0)$, but positive curvature for $T(l)$, $T(u_x)$:

  \begin{equation}
\hat{T}(\hat l)=L_0\left[1- l_c^2(\hat{l}-l_0)^2\right]\: ,
\label{TU>N}
\end{equation} 

\begin{eqnarray}
\hat{y}=y_0+\frac{|L_0|^{1/2}}{2}\left[(\hat l-l_0)\sqrt{l_c^2(\hat l-l_0)^2-1}-\log\left(2l_c(\hat l-l_0)+2\sqrt{l_c^2(\hat l -l_0)^2-1}\right)/l_c\right]. \nonumber
\label{ly>N}
\end{eqnarray}

\item Linear temperature profile $T(y)$ ($\Phi(\alpha)=0)$:

\begin{equation}
\hat{T}(\hat l)=L_{0l}(\hat l-l_0)^2\: ,
\label{TU>lin}
\end{equation} 

   \begin{eqnarray}
\hat{y}=y_0\pm\frac{L_{0l}^{1/2}}{2}\hat l(\hat l-2l_{0})\: .
\label{y>lin} \nonumber
\end{eqnarray}

\item Temperature profile $T(y)$ with positive curvature ($\Phi(\alpha)>0$):

  \begin{equation}
\hat{T}(\hat{l})=L_0\left[1+ l_c^2(\hat{l}-l_0)^2\right]\: ,
\label{TU>}
\end{equation} 

\begin{eqnarray}
\hat{y}=y_0+L_0^{1/2}\frac{1}{2 l_c}\left[\sinh^{-1}(l_c(\hat{l}-l_0))+l_c(\hat{l}-l_0)(1+c^2(\hat{l}-l_0)^2)^{1/2}\right] \: . \nonumber
\label{ly>}
\end{eqnarray}

\end{itemize}

\end{itemize}

Flow velocity profiles are obtained by making the substitution $\hat l=(\eta^*(\alpha)/\hat\gamma)\hat u_x$ in (\ref{TU<})--(\ref{TU>}). In the case of the new linear temperature profile for $\hat\Gamma(\alpha)>0$, among the two analytically possible (symmetric) solutions of the velocity profile resulting from (\ref{y>lin}), using (\ref{ugammaRed}) we find only one is physically possible, the one fulfilling $\mathrm{sg}(\partial\hat u_x/\partial\hat y)=\mathrm{sg}(\hat\gamma/\eta^*(\alpha))$. 
Note that the case $\hat\Gamma(\alpha), \Phi(\alpha)=0$ corresponds to the uniform shear flow (USF), which has been extensively studied \cite[]{Campbell}, but notice also that the USF that is not the only possible case with $\hat\Gamma(\alpha)=0$ (i.e., where viscous heating and collisional cooling are balanced). There is also the case $\hat\Gamma(\alpha)=0, \Phi(\alpha)<0$. From (\ref{y=}), the range of flow velocity for this new case is limited to $A\eta^*(\alpha)\hat{u}_x/\hat\gamma+B>0$, as $\hat{T}(\hat{y})>0$.

The solutions we describe can be applied to specific boundary conditions in a straightforward way.  As an example, consider the boundary momentum and energy balance conditions analyzed by  \cite{AN98} The boundary conditions should be expressed  in the scaled variable $\hat l$. The momentum balance conditions \cite[eq. 1 in][]{AN98} would fix the shear rate $\hat\gamma$ and the energy balance conditions \cite[eq. 2 in][]{AN98}, together with equation (\ref{TL}), would fix the constants $A, B$ (and thus, also $\Phi(\alpha)$ if equation (\ref{Phi}) is taken into account). This would determine, in the low density limit, the type of flow that would occur for given values of those boundary conditions. Either for these or other types of boundary conditions, the resulting type of profile should be described by one of the equations in (\ref{TU<})--(\ref{ly>}); i.e., our classification applies for all types of boundary conditions, and is equivalent, when restricted to symmetric boundary conditions and for the low density limit, to the description of the base steady states by \cite{AN98} and others.

It is useful to compare the sequence of steady states (\ref{TU<})--(\ref{ly>}) with the picture that emerges from prior work. For instance, in  \cite{JS83}, the change of sign of $\hat\Gamma(\alpha)$ \cite[or, analogously, $\lambda$, in the notation of][]{JS83} implies a change of sign in the curvature of $\hat T(\hat y)$ \cite[the cases of negative and positive curvature in the bulk are also referred respectively as 'sink' and 'source' walls by][see figures 1 and 2 in their work]{AN98}. However,  we see that the temperature curvature (i.e., sign of $\Phi(\alpha)$) and the sign of $\hat\Gamma(\alpha)$ do not always coincide for asymmetric boundary conditions. We would expect this to have consequences on the still unexplored stability criteria of the new steady base states we have found. Of special interest may be the asymmetric counterpart of the USF (since the USF has very extensively studied); i.e., the new case with $\hat\Gamma(\alpha)=0$,  \cite[i.e.; 'adiabatic' walls in the nomenclature of][]{AN98}, but $\Phi(\alpha)<0$, as described by equations (\ref{y=}). 

\section{Comparison with simulations and discussion
}\label{concl}

\begin{figure}
\includegraphics[height=3.85cm]{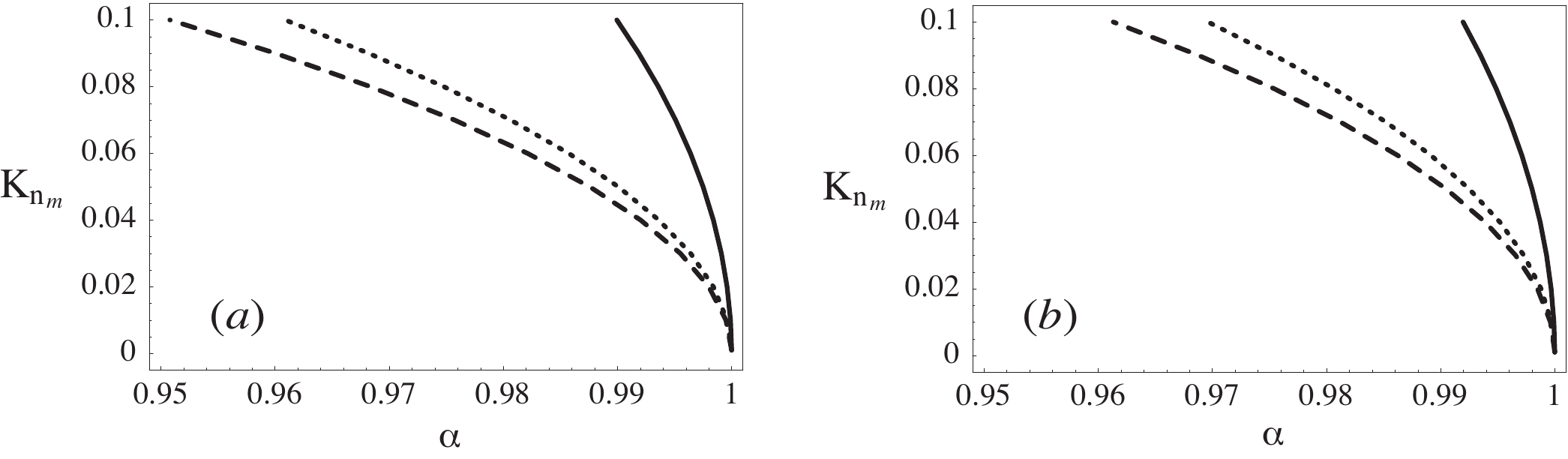}
\caption{The values $\textrm{Kn}_{-}$ (solid line), $\textrm{Kn}_{0}$ (dotted line) and $\textrm{Kn}_{+}$ (dashed line): (\textit{a}) for a 2D granular gas (disks) and (\textit{b}) for a 3D granular gas.} \label{fig1}
\end{figure}

We can calculate $\textrm{Kn}_m$, as defined in \S~\ref{redgrad}, from  (\ref{gr}, \ref{GR}). Let $\textrm{Kn}_{-}$ and $\textrm{Kn}_{+}$ stand for $\textrm{Kn}_m$ for flows with  $\hat{\Gamma}(\alpha)\le0$ and $\hat{\Gamma}(\alpha)>0$, respectively. It is also of interest to analyze $\textrm{Kn}_m$ for the case $\hat{\Gamma}(\alpha)>0$ with no shear ($\hat{\gamma}=0$), which we denote as $\textrm{Kn}_{0}$. In Fig. \ref{fig1} we show $\textrm{Kn}_m(\alpha)$ up to the transition region from Navier-Stokes to Burnett order, usually taken in the range of $\textrm{Kn}=0.1$ \cite[]{Agarwal}. These plots provide values of the minimum degree of elasticity needed to allow a Navier-Stokes regime, depending on the type of steady base state. From figure \ref{fig1} it is evident that $\textrm{Kn}_{-}$, and thus the Knudsen number for USF ($\textrm{Kn}_{-}$ indicates the value of $\textrm{Kn}$ for which $\hat\Gamma(\alpha)=0$), is always the largest of the three and the most rapidly growing. Other representative values, outside the range in the graph, are: $\textrm{Kn}_{-}=0.337$ for $\alpha=0.9$, and $\textrm{Kn}_{-}=0.527$ for $\alpha=0.7$. Nevertheless, for $\alpha>0.995$ the system has  $\textrm{Kn}_{-}$  $\ll1$, and thus, there are uniform shear flows that may be described by the Navier-Stokes hydrodynamics. Another interesting result in Fig. \ref{fig1} is that the sheared states are less restrictive compared to the non-sheared ones (as $\textrm{Kn}_{0}>\textrm{Kn}_{+}$, systematically). Also, we see that Navier-Stokes hydrodynamics does not guarantee a correct description of steady shear flows in the granular gas for $\alpha<0.95$ (although this does not mean that NS description of the base steady states will completely fail for $\alpha<0.95$). 

\begin{figure}
\begin{center}
\includegraphics[height=3.8cm]{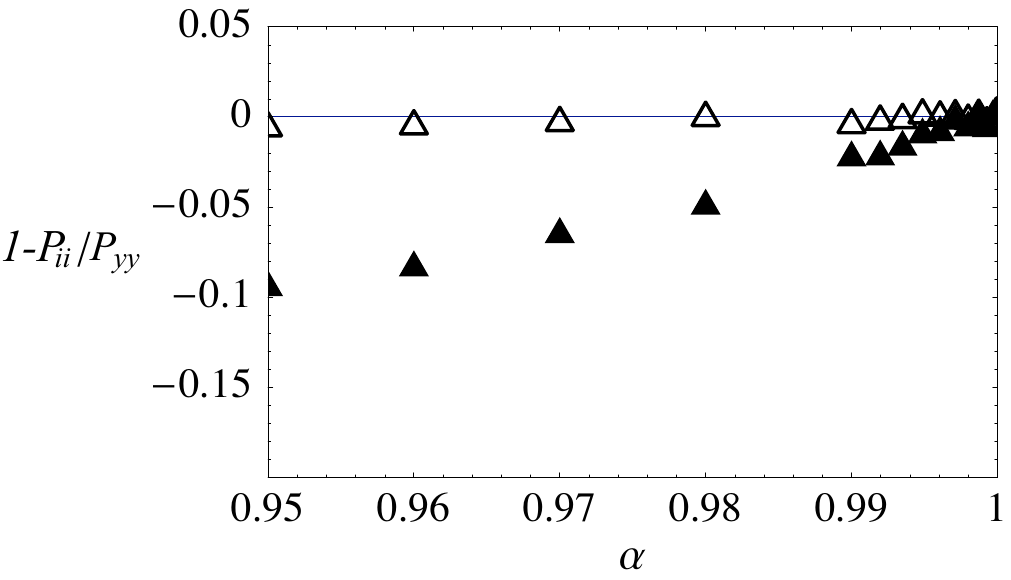}
\end{center}
\caption{The reduced normal stress differences $1-P_{xx}/P_{yy} (\blacktriangle), 1-P_{zz}/P_{yy} (\vartriangle$) for uniform shear flow ($\hat{\Gamma}(\alpha)=0, A=0, B\neq0$) in a 3D granular gas are almost zero for $\alpha>0.995$, approaching the Navier-Stokes regime. DSMC simulations are performed in the dilute limit.} \label{fig2}
\end{figure}

In order to qualitatively asses the reliability of the interpretation of results in Fig. \ref{fig1}, we have performed computer simulations of the USF based on the direct simulation Monte Carlo (DSMC) method \cite[as implemented by][]{DUSF}. Concretely, we look at the normal stress differences, a non-linear effect (beyond NS) that appears in the problem for high enough Knudsen number \cite[]{GoldJFM}. We plot in figure \ref{fig2} the values of $1-P_{ii}/P_{yy}$ (with $i=x, z$) for the USF in a low density granular gas. The values of $1-P_{ii}/P_{yy}$ are very close to zero (within $1~\%$) for $\alpha>0.995$ but $1-P_{xx}/P_{yy}$ starts increasing significantly (i.e. at least a departure from NS in one aspect is observed) for $\alpha<0.995$ (see \cite{DUSF} for more simulation data in a smooth hard sphere granular gas  and \cite{Campbell} for a rough sphere gas).
The results in figure \ref{fig2} are in agreement with the range for which $\textrm{Kn}_{-}$ (figure \ref{fig1}) becomes non-negligible.

A more severe test is to compare directly the hydrodynamic steady profiles resulting from DSMC and MD simulations with the theoretical predictions of our study, particularly for non-homogeneous states. 
We have carried out DSMC simulations of hard spheres as implemented  by \cite{VGS08} in their DSMC results for the Boltzmann equation. The MD simulations in this work use a standard event-driven algorithm with the hard sphere collisional model \cite[more details on methodology can be found elsewhere, for instance in the book by][]{Rapaport}. In both MD and DSMC simulations we use 'stochastic' boundary conditions \cite[]{Rapaport}; i.e., the heating walls act by changing the colliding particle velocity components so that new random values are assigned to them according to gaussian (for the components parallel to the wall) and Rayleigh (for the normal component) distribution functions with the same temperature as the pertaining wall \cite[this type boundary conditions are commonly used in the field, see for instance][]{GHW07}. 
If there is shear, an $x$-component constant value, characteristic of each wall, is added to the velocity of the colliding particle \cite[]{VGS08}. 
Thus, the input for the boundary conditions in the simulations are the fluid temperature and flow velocity next to the walls, and $\hat\gamma$, $\hat\Gamma(\alpha)$, and $\Phi(\alpha)$ are extracted from steady state simulation data, with the aid of relations defined in \S\ref{2D} \footnote{If the simulation/experimental $\hat T(\hat y)$ and $\hat u(\hat y)$ are measured then: $\hat\Gamma_{exp}=(\partial^2\hat T/\partial \hat u_x^2)(\hat\gamma/\eta^*)_{exp}^2$, with $(\hat\gamma/\eta^*)_{exp}=\hat T^{1/2} \partial \hat u_x/\partial\hat y$.}. The simulations are performed in dilute systems (a packing fraction $\nu\sim7\times 10^{-3}$ is used in MD simulations) and the number of particles is of the order of $N=2\times10^6$ for DSMC and $N=40000$ for MD.

\begin{figure}
\begin{center}
\includegraphics[height=4.30cm]{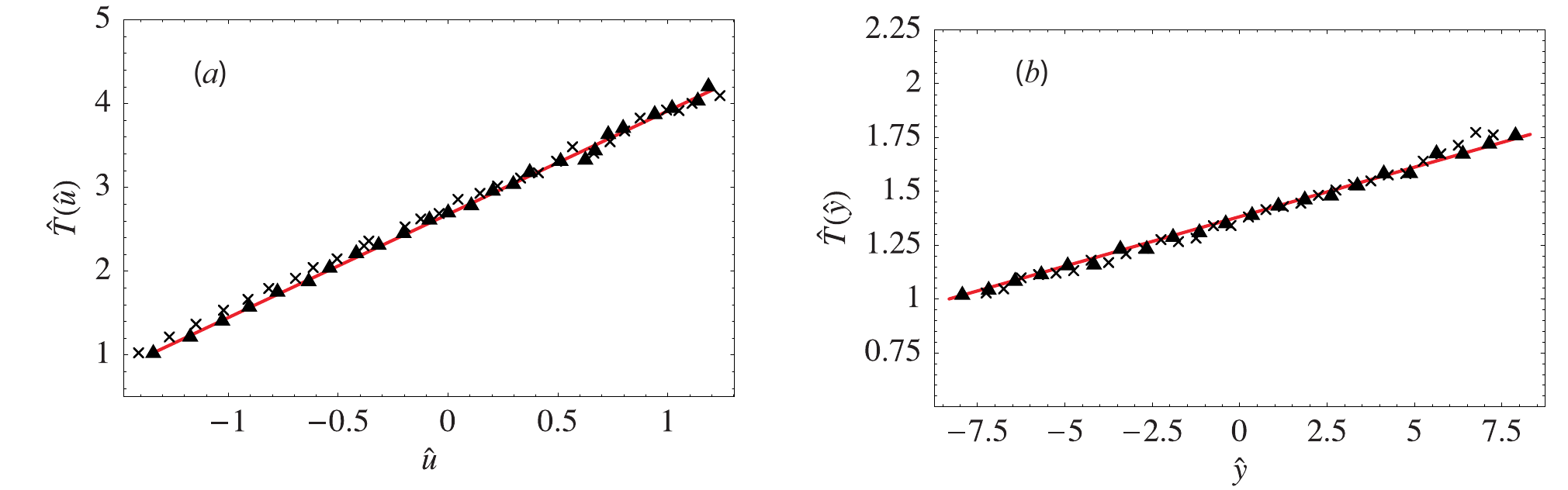}
\end{center}
\caption{Comparison between theoretical (line), DSMC ($\blacktriangle$) and MD simulations ($\times$) temperature profiles for two new steady states. (\textit{a}) A state of the type $\hat\Gamma(\alpha)=0$ like USF but with $\Phi<0$, for which theoretically $\hat T(\hat l)$, $\hat T(\hat u)$ are non-uniform and linear. (\textit{b}) a state with $\hat\Gamma(\alpha)>0$ and $\Phi=0$ and for which theoretically $\hat T(\hat y)$ is linear. In both cases $\alpha=0.99$, $d=3$. In (\textit{a}) the Navier-Stokes theory predicts $\hat\gamma= 0.117$, which compares (for $A=0.137$) well with simulations, from which we measured $\hat\gamma= 0.114$, $A=0.149$ in DSMC ($2.6\%$ and $8.6\%$ difference respectively; $A$ values are here extracted from the slope of $\hat T(\hat u)$, see equations (\ref{y=}) and (\ref{ul})) and $\hat\gamma= 0.122$, $A=0.140$ in MD ($4.6\%$ and $4.1\%$ difference respectively). In (\textit{b}),  with $\hat\gamma=0.091$, the theoretical prediction for $\hat\Gamma(\alpha)$ is  $\hat\Gamma=1.101\times 10^{-3}$, which compares well with the DSMC value  $\hat\Gamma=1.12\times 10^{-3}$ ($1.7\%$ difference) and the MD value $\hat\Gamma=1.03\times 10^{-3}$ ($6.4\%$ difference).}\label{figlineal}
\end{figure}

We discuss first the new steady states  with linear temperature profile: linear $\hat T(\hat u)$ (i.e., with $\hat\Gamma (\alpha)=0$ and $\Phi(\alpha)<0$) and linear $\hat T (\hat y)$ (i.e., with $\hat\Gamma (\alpha)>0$ and $\Phi(\alpha)=0$). Figure \ref{figlineal} shows the results for $\alpha=0.99$ ($\alpha$ sufficiently high to allow Navier-Stokes hydrodynamics, see figure \ref{fig1}). The agreement between the theoretical and the DSMC profiles is excellent for both cases. The deviation from a linear profile for DSMC is just $0.54\%$ and $0.73\%$ respectively (and the MD data are similarly close to linearity as well, see figure \ref{figlineal} caption for more details).  We have performed a series of simulations with decreasing shear rate and have found all the profiles corresponding to the theoretical solutions (\ref{TU<})--(\ref{TU>}) (see supplementary material). For $\hat\Gamma(\alpha)<0$, a state with $\Phi(\alpha)<0$ is found. Decreasing the shear rate, we reach a flow with $\hat\Gamma (\alpha)=0, \Phi(\alpha)<0$ (or the USF, with $\hat\Gamma (\alpha)=0, \Phi(\alpha)=0$ if the fluid near both walls is at the same temperature). If we further decrease shear rate we find states with $\hat\Gamma (\alpha)>0$ with, consecutively, $\Phi(\alpha)<0, \Phi(\alpha)=0, \Phi(\alpha)>0$ (or only $\Phi(\alpha)>0$ if the fluid near both walls is at the same temperature)\footnote{We refer the reader to the supplementary material accompanying this article for figures of all different types of flows, as extracted from  (\ref{TU<})--(\ref{TU>}).}. Furthermore, we found the same sequence of steady states for simulations far from the Navier-Stokes order, which is an indication that the qualitative behavior described here probably persists  beyond the range of small $\textrm{Kn}$. This is perhaps not surprising since \cite{JStat} have shown that for a granular gas \cite[and also, for a granular impurity, see][]{VGS08}, the Couette flows obey hydrodynamic equations that share the same structure with those at Navier-Stokes order, albeit with new nonlinear transport coefficients and $\hat\Gamma (\alpha)$, dependent on a constant shear rate, that would have to be calculated \cite[]{JStat, ExUnifAn, VGS08}. 

\begin{figure}
\begin{center}
\includegraphics[height=4.0cm]{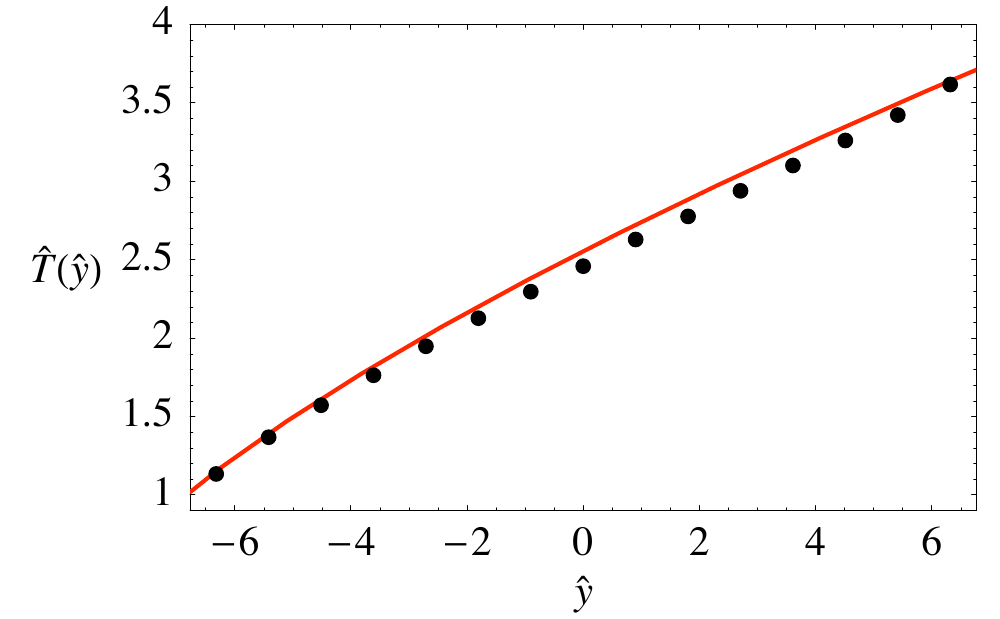}
\end{center}
\caption{Temperature profile with negative curvature (elastic gas-like) for a granular gas ($\alpha=0.999$) heated from the walls and no shear ($\hat\gamma=0$). This type of profile has been not reported before in granular gases. The line corresponds to the theoretical profile with $\Phi=-0.0321$ and points correspond to MD simulation data, with $\Phi=-0.0318$ ($1\%$ difference; the theoretical profile was adjusted to make the temperature boundary values coincide, and $\Phi$ in the simulation is obtained using the relation (\ref{ecder})). }\label{figphineg}
\end{figure}

MD simulations results in figure \ref{figphineg} show another type of profile not reported (to our knowledge) in the literature on granular gases in contact with thermal walls. In this case there is no shear in the system ($\hat\gamma=0$) and the temperature profile shows the negative curvature characteristic of an elastic gas subjected to two thermal walls at different temperatures. As in the linear temperature profiles, there is good agreement between simulation and theory. The boundary conditions we used for the MD simulations are very common in the literature \cite[]{GHW07}, but one could use in the MD code boundary conditions closer to experimental boundary conditions; for instance,  bumpy moving walls to produce shear \cite[]{L96}, or vertically vibrating walls to heat the granular gas \cite[]{VU08}. It is reasonable to expect  good agreement between our theory and the simulation results with these more realistic boundary conditions, since it has been found that this agreement exists in experiments \cite*[]{YHC02} for the steady states derived previously.
 
The physical origin of the sometimes surprising behavior of the temperature profile is the temperature dependence of the hydrodynamic transport coefficients, combined with the effects of constant pressure. Consider, for example, the case of a linear temperature profile when collisional cooling exceeds viscous heating.  If the transport coefficients were approximately constant, the heat flux would be constant across the system, and there would be no source to replace the energy lost to collisional cooling.  However, in the hydrodynamic equations the heat flux is actually proportional to $\sqrt{T}dT/dy$, and therefore is not in general constant for a linear temperature profile. In the case of the linear profile, the decrease in heat flux from the hot side of the system to the cold side compensates for the collisional cooling.  A perhaps surprising result is that if this balance is satisfied at one point in the flow, it is satisfied everywhere (or, more generally, that the curvature of the temperature profile is constant, even though the temperature and density are not).  This is because the hydrodynamic fields scale in the same way as the pressure, which is constant across the system.

The results for DSMC and MD simulations supply information at two different levels. DSMC results show that our theory, resulting from an expansion of the solution of the kinetic equation, approaches in effect the exact solution of the kinetic equation (yielded by DSMC). MD simulations help to check that the description resulting from the kinetic equation approaches a 'real' system.  The states that we have described in molecular dynamics simulations should be observable, in principle, in experiments. The temperature and/or heat flux of the sheared granular gas at the boundary is determined by the physical properties of the shearing walls  (roughness, friction, vibration, etc.). Starting, for example, from symmetric relatively smooth non-vibrating walls, and then systematically increasing the roughness and/or vibration amplitude of one wall, the flow states we have identified can be explored.  

The unidimensional steady states analyzed here are the starting point for more complex studies on granular fluid flow \cite[]{Batchelor}. 
For example, the fact that the linear $\hat T(\hat u)$  shows exactly the same transport properties and rheology as the USF in the regime beyond Navier-Stokes order \cite[]{VGS08,SVG09}, will allow the measurement of heat flux transport coefficients directly from a numerical solution of the Boltzmann equation in steady state, something that was not possible previously \cite[]{G06}. Preliminary DSMC results confirm that this state also appears beyond NS order \footnote{DSMC data showing the sequence of steady states appearing for varying shear rate, as described above, and linear $\hat T(\hat  u)$ for NS order and beyond, may be found in the website http://www.unex.es/fisteor/$\sim$fvega/dsmc.html. Our DSMC codes are also available to the reader, please contact fvega@unex.es.} and that the (non-linear) shear viscosity values (from DSMC) for both the USF and the new states show nearly identical values \cite*[]{SVG09}. 
Another possible application on which we are working is the Brazil-nut segregation problem, since the input of the solutions of the profiles is needed in order to determine the segregation criteria \cite*[]{GV08}. Additionally, results in figure \ref{figphineg} show that the positive curvature caused by collisional cooling can be reversed if the thermal walls are at sufficiently different temperatures. Thus, the results allow the extension of the study of deviations from NS hydrodynamics for a heated granular gas \cite*[]{HGW08} (no flow velocity) to the three classes of steady states that will appear, depending on the wall temperature difference, as described by (\ref{TU>N})--(\ref{ly>}). It would also be interesting to study the effect that this heat flow gradient sign reversal mechanism would have on hydrodynamic stability problems applied to granular gases, such as the buoyancy instability mechanism in a fluid layer between two walls at different temperatures (Rayleigh-B\'enard convection).


\begin{acknowledgments}
We thank Andr\'es Santos and Vicente Garz\'o for useful comments on the early version of this manuscript, and A. E. Lobkovsky for his event-driven simulation code. This work was supported by NASA under award number NNC04GA63G. Also, F. V. R. acknowledges financial support from the Spanish Science Ministry as an FPI-SEEU Fellow (ref. GT-2002-0023) and through "Juan de la Cierva" research program.  
\end{acknowledgments}

\bibliographystyle{jfm}
\bibliography{GFT09}

\vspace{3 mm}

\end{document}